# Equations of Motion that Recognize Biochemical Patterns

## Eisuke Chikayama*

*Advanced NMR metabomics Research Unit, Metabolome Research Group, RIKEN Plant Science Center, 1-7-22 Suehiro-cho, Tsurumi-ku, Yokohama, Kanagawa 230-0045, Japan*

* Author to whom correspondence should be addressed. Tel.: +81-45-503-9490. Fax: +81-45-503-9489. E-mail: chika@psc.riken.jp.

### Abstract
Equations of motion that recognize biochemical patterns are described. The equations are partial differential equations in a continuous multiple component system in which adequate initial and boundary conditions are given. The biochemical patterns are spatiotemporal distributions of multiple biochemical components that can be regarded as a continuum in concentration and mass flux. Recognizing biochemical patterns lead to a universal property of the equations, which is also mathematically demonstrated, that the devised equations are sufficient to approximate any orbits in an arbitrary dynamical system, even though that of expectedly seen in complex biological systems. This theory can be applied to also a non-biological system that can be regarded as a continuum comprised of any multiple components such as liquids, solids, and nonlinear viscoelastic materials.

### Theoretical background

Natural materials are approximated with a continuum consist of multiple components. Our theory is for such a multiple component system comprised of an arbitrary number of nonlinear viscoelastic solids and viscoelastic liquids. Such a system can be strictly described by elemental equations of continuum mechanics [1]:

$$\frac{\partial X_i}{\partial t} = R_i - \nabla \cdot \vec{N}_i, \qquad (1)$$

$$\frac{\partial \vec{N}_i}{\partial t} = \vec{G}_i - \nabla \cdot \hat{\Phi}_i - \nabla \cdot \left(\frac{\vec{N}_i \vec{N}_i}{X_i}\right), \qquad (2)$$

where, $X_i$ is the concentration of the $i$-th component (a scalar), $\vec{N}_i$ is the mass flux of the $i$-th component (a vector), $R_i$ is the rate of generation in concentration of the $i$-th component (so-called reaction term, a scalar), $\vec{G}_i$ is the exerted body force for the $i$-th component (a vector), $\hat{\Phi}_i$ is the exerted stress for the $i$-th component (a tensor), and derivations are performed with respect to $x$, $y$, $z$, or $t$, space and time coordinates, respectively. The second term in (1) and the third term in (2) are the net rates of addition of mass per unit volume and mass flux, respectively, by transport. In general, $R_i$, $\vec{G}_i$, and $\hat{\Phi}_i$ are nonlinear with respect to physical state variables. The equations (1) and (2) correspond to the law of conservation of mass and equations of motion, respectively, in the continuum.

In our theory, fractional statistics [2] is essential. Fractional statistics has been proposed aiming at the unification of the three elemental statistics theories, i.e., Fermi-Dirac, Boltzmann, and Bose-Einstein statistics. A probability distribution function in the fractional statistics scheme is then simply described as,

$$p(E) = \frac{1}{\exp(\beta(E-\mu)) + \delta}, \qquad (3)$$

where $p$ is the probability density of the system having an energy, $E$; $\beta$ represents the reciprocal of temperature; $\mu$ is the chemical potential; and $\delta$ is a real constant specifying a type of distribution. $\delta = 1, 0, -1$ represent Fermi-Dirac, Boltzmann, and Bose-Einstein distribution, respectively.

### Proposed equations of motion for biochemical patterns

In our system, as in the elastic theory of materials [3], it is supposed that the system has a memory of past orbits, i.e., at any point $(x_1, y_1, z_1)$ in the system and the time $t$, $\partial X_i/\partial t(x_1, y_1, z_1, t)$ or $\partial \vec{N}_i/\partial t(x_1, y_1, z_1, t)$, the left-hand sides of (1) or (2), can depend on any numbers of combinations of $j$-th spatiotemporal distributions, $X_j(x, y, z, t-\tau)$ and $\vec{N}_j(x, y, z, t-\tau)$; where $\tau$ is a parameter designating a time to the past from the current. This means that the given spatiotemporal (, only the past if $i = j$,) distributions of multiple components of $X_j$ and $\vec{N}_j$ uniquely define evolution of the system at the time $t$, the right-hand sides of (1) and (2).

Proposed equations of motion for biochemical patterns,

$$\frac{\partial X_i}{\partial t} = \sum_j a_j s_j - \nabla \cdot \vec{N}_i \qquad (4)$$

$$\frac{\partial \vec{N}_i}{\partial t} = \sum_j b_j \vec{s}_j - \sum_k c_k \nabla \cdot \hat{s}_k - \nabla \cdot \left(\frac{\vec{N}_i \vec{N}_i}{X_i}\right) \qquad (5)$$

$$\vec{s} = (s_x, s_y, s_z) \qquad (6)$$

$$\hat{s} = \begin{pmatrix} s_{xx} & s_{xy} & s_{xz} \\ s_{yx} & s_{yy} & s_{yz} \\ s_{zx} & s_{zy} & s_{zz} \end{pmatrix}, \qquad (7)$$

are derived by using the expansion of the non-linear terms, $R_i$, $\vec{G}_i$, and $\hat{\Phi}_i$ in equations (1) and (2), to series of fractional statistics functions. In the equations (4) and (5), $a_j$, $b_j$, and $c_k$ are real constants, and each of expanded series is a linear combination of

$$s(Y) = \prod_{j=1}^{M} \frac{1}{\exp(\beta_j(Z_j[Y] - \mu_j)) + \delta_j},$$
$$(1 \leq M \leq \infty) \qquad (8)$$

a direct product of fractional statistics functions, which results in a real value; where $\beta_j$, $\mu_j$, and $\delta_j$ are real constants; $Z_j[Y]$ is a real-valued scalar field and a functional, which can be nonlinear, of $Y$; $Y$ is a function space comprised of all the $X_j(x, y, z, t-\tau)$ and $\vec{N}_j(x, y, z, t-\tau)$ that are physically feasible in the concerned system. Note that our theory only adopts the mathematical form of fractional statistics functions, (3), and there is no thermodynamical meaning.

We have mathematically demonstrated that our equations (4) and (5), which are derived from expanded series of fractional statistics functions, is sufficient to approximate (1) and (2), a very broad class of nonlinear infinite dynamical systems. It is described elsewhere.